\documentclass[a4paper]{amsart}

\usepackage{amsmath}
\usepackage{amssymb}
\usepackage{amsthm}
\usepackage{bbm}

\newtheorem{theorem}{Theorem}

\newcommand{\expec}{\mathbf{E}}

\newcommand{\e}{\mathrm{e}}
\newcommand{\dd}{\mathrm{d}}

\title{Moment Explosion in the LIBOR Market Model}

\author{Stefan Gerhold}

\address{Vienna University of Technology, Wiedner Hauptstra\ss{}e 8--10,
A-1040 Vienna, Austria}

\email{sgerhold at fam.tuwien.ac.at}

\date{\today}

\begin{document}

\begin{abstract}
  In the LIBOR market model, forward interest rates are log-normal under
  their respective forward measures. This note shows that their distributions under
  the other forward measures of the tenor structure have approximately
  log-normal tails.
\end{abstract}




\maketitle

\section{Introduction}

The LIBOR market model~\cite{BrGaMu97} is one of the most popular models
for pricing and hedging interest rate derivatives.
Its state variables are forward interest rates $F_n(t):=F(t;T_{n-1},T_n)$,
spanning time periods $[T_{n-1},T_n]$, where
\[
  0<T_0 <T_1< \dots <T_M
\]
is a fixed tenor structure.
Under the $T_M$-forward measure~$\mathbb{Q}^M$, which has as numeraire the zero coupon bond
maturing at~$T_M$, the dynamics of the forward rates are 
\begin{align*}
  \dd F_n(t) &= -\sigma_n(t) F_n(t) \sum_{j=n+1}^M \frac{\rho_{nj}\tau_j\sigma_j(t)F_j(t)}
    {1+\tau_j F_j(t)}\dd t + \sigma_n(t) F_n(t) \dd W_n(t),\\
    &\quad 1\leq n < M,\\
  \dd F_M(t) &= \sigma_M(t) F_M(t) \dd W_M(t).
\end{align*}
Here, $\sigma_n$ are some positive deterministic volatility functions, and~$W$
is a vector of standard Brownian motions with instantaneous correlations
$\dd W_i(t) \dd W_j(t) = \rho_{ij} \dd t$. Moreover, $\tau_n=\tau(T_{n-1},T_n)$
denotes the year fraction between the tenor dates~$T_{n-1}$ and~$T_n$.

Note that each rate~$F_n$ is a geometric Brownian Motion under
its own forward measure, while it has
a non-zero drift under the other forward measures.
A popular approximation of the above dynamics is obtained by ``freezing the drift'':
\begin{align*}
  \dd F_n^{\mathrm{fd}}(t) &= -\sigma_n(t) F_n^{\mathrm{fd}}(t) \sum_{j=n+1}^M \frac{\rho_{nj}\tau_j\sigma_j(t)F_j(0)}
    {1+\tau_j F_j(0)}\dd t + \sigma_n(t) F_n^{\mathrm{fd}}(t) \dd W_n(t),\\
    &\quad 1\leq n < M,\\
  \dd F_M^{\mathrm{fd}}(t) &= \sigma_M(t) F_M^{\mathrm{fd}}(t) \dd W_M(t).
\end{align*}
Since the drifts are now deterministic, the new rates~$F_n^{\mathrm{fd}}$
are just geometric Brownian motions,
which allows for explicit pricing formulas for many interest-linked products.
As a piece of evidence for the quality of this approximation, we show
in the present note that, for fixed $t>0$, the distribution of~$F_n^{\mathrm{fd}}(t)$
has roughly the same tail heaviness as the distribution of~$F_n(t)$.

\section{Main Result}

If~$X$ is any log-normal random variable, so that $\log X\sim \mathcal{N}(\mu,\sigma^2)$
for some real~$\mu$ and positive~$\sigma$, then
\begin{equation}\label{eq:log moment}
  \sup\{v : \expec[\e^{v \log^2 X}] < \infty \}  = \frac{1}{2\sigma^2}.
\end{equation}
This follows from
\[
  \expec[\e^{v\log^2 X}] = \frac{1}{\sqrt{1-2\sigma^2 v}} \exp \left(
  \frac{\mu^2 v}{1-2\sigma^2 v}\right), \qquad v < \frac{1}{2\sigma^2}.
\]
Our main result shows that~$F_n(t)$ has approximately log-normal tails, in the sense
that the left-hand side of~\eqref{eq:log moment} is finite and positive
if~$X$ is replaced by~$F_n(t)$. Furthermore, this ``critical moment''
is the same for~$F_n(t)$ and the frozen drift approximation~$F_n^{\mathrm{fd}}(t)$.
\begin{theorem}\label{thm:main}
  In the log-normal LIBOR market model, we have for
  all $t>0$ and all $1\leq n\leq M$
  \begin{align*}
    \sup\{v : \expec^{M}[\e^{v \log^2(F_n(t))}] < \infty \}  &=
    \sup\{v : \expec^{M}[\e^{v \log^2(F_n^{\mathrm{fd}}(t))}] < \infty \} \\
    &=  \frac{1}{2\int_0^t \sigma_n(s)^2 \dd s}.
  \end{align*}
\end{theorem}
\begin{proof}
  Note that the latter equality is obvious from~\eqref{eq:log moment},
  since~$F_n^{\mathrm{fd}}(t)$ is log-normal
  with log-variance parameter $\sigma^2=\int_0^t \sigma_n(s)^2 \dd s$.
  We now show the first equality.
  Recall that the measure change from the $T_{n}$-forward measure to the
  $T_{n-1}$-forward measure is effected by the likelihood
  process~\cite{Bj04}
  \[
    \left. \frac{\dd \mathbb{Q}^{n}}{\dd \mathbb{Q}^{n-1}} \right|_{\mathcal{F}_t}
      = \frac{1+\tau_n F_n(0)}{1+\tau_n F_n(t)}.
  \]
  Therefore, putting $\phi(x)=\exp(\log^2 x)$, we obtain
  \begin{align*}
    \expec^M[\phi(F_n(t))^v] &= \expec^{M-1}\left[\phi(F_n(t))^v \times
      \frac{1+\tau_M F_M(0)}{1+\tau_M F_M(t)} \right] \\
    &= \dots = \\
    &= \expec^{n}\left[\phi(F_n(t))^v \prod_{i=n+1}^M \frac{1+\tau_i F_i(0)}{1+\tau_i F_i(t)} \right] \\
    &\leq  \expec^{n}[\phi(F_n(t))^v] \prod_{i=n+1}^M (1+\tau_i F_i(0)),
  \end{align*}
  hence
  \[
    \sup\{v : \expec^{n}[\phi(F_n(t))^v] < \infty \}
      \leq \sup\{v : \expec^{M}[\phi(F_n(t))^v] < \infty \}.
  \]
  On the other hand, for $1< k \leq M$ we have
  \begin{align*}
    \expec^{k-1}[\phi(F_n(t))^v] &=
      \expec^{k}\left[\phi(F_n(t))^v \times \frac{1+\tau_{k} F_{k}(t)}{1+\tau_{k} F_{k}(0)} \right] \\
    &= \frac{1}{1+\tau_{k} F_{k}(0)} \left( \expec^{k}[\phi(F_n(t))^v]
      + \tau_{k} \expec^{k}[F_{k}(t) \phi(F_n(t))^v] \right). 
  \end{align*}
  Now let $\varepsilon>0$ be arbitrary, and define~$q$ by $\frac1q + \frac{1}{1+\varepsilon}=1$.
  Then H{\"o}lder's inequality yields
  \[
    \expec^{k}[F_{k}(t) \phi(F_n(t))^v] \leq \expec^{k}[F_{k}(t)^q]^{1/q}
      \times \expec^{k}[ \phi(F_n(t))^{v(1+\varepsilon)}]^{1/(1+\varepsilon)}.
  \]
  By the finite moment assumption, we obtain the implication
  \[
    \expec^{k}[ \phi(F_n(t))^{v(1+\varepsilon)}] < \infty \quad
      \Longrightarrow \quad \expec^{k-1}[ \phi(F_n(t))^v] < \infty, \qquad v\in\mathbb{R}.
  \]
  (Note that the left-hand side implies $\expec^{k}[ \phi(F_n(t))^{v}] < \infty$.)
  
  Inductively, this leads to the implication
  \[
    \expec^{M}[ \phi(F_n(t))^{v(1+\varepsilon)^{M-n}}] < \infty \quad
      \Longrightarrow \quad \expec^{n}[ \phi(F_n(t))^v] < \infty, \qquad v\in\mathbb{R}.
  \]
  Therefore, we find
  \begin{align*}
    \sup\{v : \expec^{n}[\phi(F_n(t))^v] < \infty \}
      &\geq \sup\{v : \expec^{M}[ \phi(F_n(t))^{v(1+\varepsilon)^{M-n}}] < \infty \} \\
    &= \frac{1}{(1+\varepsilon)^{M-n}} \sup\{v : \expec^{M}[ \phi(F_n(t))^{v}] < \infty \}.
  \end{align*}
  Since~$\varepsilon$ was arbitrary, 
  \[
    \sup\{v : \expec^{n}[\phi(F_n(t))^v] < \infty \} \geq
       \sup\{v : \expec^{M}[\phi(F_n(t))^v] < \infty \}
  \]
  follows, which finishes the proof.
\end{proof}

\bibliographystyle{siam}
\bibliography{../gerhold}

\end{document}